\documentclass[conference]{IEEEtran}
\IEEEoverridecommandlockouts

\usepackage{cite}
\usepackage{amsmath,amssymb,amsfonts}
\usepackage{algorithmic}
\usepackage[linesnumbered,ruled]{algorithm2e}
\usepackage{amsthm}
\usepackage{array}
\usepackage{graphicx}
\usepackage{textcomp}
\usepackage{paralist}
\usepackage{subfigure}
\usepackage{hyperref}
\usepackage{multirow}
\usepackage{url}
\usepackage{setspace}
\usepackage{tabu}
\usepackage{tablefootnote}
\usepackage{romannum}
\usepackage{enumitem}
\usepackage{arydshln}
\usepackage[normalem]{ulem}
\usepackage{threeparttable}
\usepackage{url}
\usepackage{xcolor}
\usepackage{xspace}
\usepackage{soul}
\usepackage{booktabs}
\usepackage{tabularx}
\usepackage[most]{tcolorbox}
\usepackage{multirow}

\definecolor{background}{rgb}{0.94, 0.97, 1.0}
\definecolor{edge}{rgb}{0.32, 0.48, 0.72}
\newtcolorbox{mybox}{colback=background!50,%gray background
colframe=edge,% black frame colour
width=\columnwidth,% total width
arc=2mm, 
auto 
outer 
arc
}

\newcommand{\qoa}{{QoA}\xspace}
\newcommand{\company}{Huawei Cloud\xspace}

\begin{document}
\title{Characterizing and Mitigating Anti-patterns of Alerts in Industrial Cloud Systems}


% \author{Anonymous Author(s)}
% \author{Submission Category: Practical Experience Report}

\author{
  \IEEEauthorblockN{
    Tianyi Yang\IEEEauthorrefmark{1},
    Jiacheng Shen\IEEEauthorrefmark{1},
    Yuxin Su\IEEEauthorrefmark{2}\thanks{Yuxin Su is the corresponding author.},
    Xiaoxue Ren\IEEEauthorrefmark{1},
    Yongqiang Yang\IEEEauthorrefmark{3}, and
    Michael R. Lyu\IEEEauthorrefmark{1}
  }

  \IEEEauthorblockA{\IEEEauthorrefmark{1}Department of Computer Science and Engineering, The Chinese University of Hong Kong, Hong Kong, China.\\
    Email: \{tyyang, jcshen, lyu\}@cse.cuhk.edu.hk; xiaoxueren@cuhk.edu.hk}

  \IEEEauthorblockA{\IEEEauthorrefmark{2}School of Software Engineering, Sun Yat-Sen Univeristy, Zhuhai, China. Email: suyx35@mail.sysu.edu.cn}

  \IEEEauthorblockA{\IEEEauthorrefmark{3}Computing and Networking Innovation Lab, Cloud BU, Huawei, Shenzhen, China. Email: yangyongqiang@huawei.com}
}


\maketitle

% add page numbers (but IEEE template do not have page numbers!)
% \pagestyle{plain}

\begin{abstract}

    Alerts are crucial for requesting prompt human intervention upon cloud anomalies.
    The quality of alerts significantly affects the cloud reliability and the cloud provider's business revenue.
    In practice, we observe on-call engineers being hindered from quickly locating and fixing faulty cloud services because of the vast existence of misleading, non-informative, non-actionable alerts.
    We call the ineffectiveness of alerts ``anti-patterns of alerts''.
    To better understand the anti-patterns of alerts and provide actionable measures to mitigate anti-patterns, in this paper, we conduct the first empirical study on the practices of mitigating anti-patterns of alerts in an industrial cloud system.
    We study the alert strategies and the alert processing procedure at \company, a leading cloud provider.
    Our study combines the quantitative analysis of millions of alerts in two years and a survey with eighteen experienced engineers.
    As a result, we summarized four individual anti-patterns and two collective anti-patterns of alerts.
    We also summarize four current reactions to mitigate the anti-patterns of alerts, and the general preventative guidelines for the configuration of alert strategy.
    Lastly, we propose to explore the automatic evaluation of the Quality of Alerts (\qoa), including the indicativeness, precision, and handleability of alerts, as a future research direction that assists in the automatic detection of alerts' anti-patterns.
    The findings of our study are valuable for optimizing cloud monitoring systems and improving the reliability of cloud services.

\end{abstract}


\section{Introduction}
\label{sec:intro}

The boost of cloud adoption puts forward higher requirements on the reliability and availability of cloud services.
Typically, cloud services are organized and managed as microservices that interact with each other and serve user requests as a whole.
In a large-scale cloud microservice system, unplanned microservice anomalies happen from time to time.
Some anomalies are transient, while others persist and require human intervention.
If anomalies are not detected and mitigated timely, they may cause severe cloud failures and incidents, affect the availability of cloud services, and deteriorate user satisfaction~\cite{DBLP:conf/sigsoft/ChenKLZZXZYSXDG20}.
Hence, prompt detection, human intervention, and mitigation of service anomalies are critical for the reliability of cloud services.
To accomplish that, cloud service providers employ large-scale cloud monitoring systems that monitor the system state and generate alerts that require human intervention.
Whenever anomalous states of services emerge, alerts will be generated to notify engineers to prevent service failures.

In a cloud system, an \textbf{alert} is a notification sent to On-Call Engineers (OCEs), of the form defined by the \textit{alert strategy}, of a specific abnormal state of the cloud service, i.e., an \textbf{anomaly}.
A severe enough alert (or a group of related alerts) can escalate to an \textbf{incident}, which, by definition, is any unplanned interruption or performance degradation of a service or product, which can lead to service shortages at all service levels~\cite{DBLP:conf/sigsoft/ChenKLZZXZYSXDG20}.
An \textbf{alert strategy} defines the policy of alert generation, i.e., \textit{when to generate an alert}, \textit{what attributes and descriptions an alert should have}, and \textit{to whom the alert should be sent}.
Once an OCE receives an alert, the OCE will follow the corresponding predefined Standard Operating Procedure (\textbf{SOP}) to inspect the state of the cloud service and mitigate the service anomaly based on their domain knowledge.
% With the advancement of AIOps~\cite{DBLP:conf/icsoc/NotaroCG20}, some SOPs can be automatically conducted.
The \textit{alert strategies} and \textit{SOPs} are two key aspects to ensure a prompt and effective response to cloud alerts and incidents.
In industrial practice, the two aspects are often considered and managed together because improperly designed alert strategies may lead to non-informative or delayed alerts, affecting the diagnosis and mitigation of the cloud alerts and incidents.
We call the unified management of \textit{alert strategies} and \textit{SOPs} \textbf{alert governance}.
Table~\ref{tab:terminology} summarizes the terminologies used in this paper.

\begin{table*}[t]
  \centering
  \small % font size
  \caption{The Terminology Adopted in This Paper.}
  \label{tab:terminology}
  \begin{tabularx}{2\columnwidth}{lX}
    \toprule
    \textbf{Term}             & Explanation                                                                                                                                                                                                     \\
    \midrule
    \textbf{Anomaly}          & A deviation from the normal state of the cloud system, which will possibly trigger an alert.                                                                                                                    \\
    \textbf{Alert}            & A notification sent to On-Call Engineers (OCEs), of the form defined by the alert strategy, of a specific anomaly of the cloud system.                                                                          \\
    \textbf{Incident}         & Any unplanned interruption or performance degradation of a service or product, which can lead to service shortages at all service levels~\cite{DBLP:conf/sigsoft/ChenKLZZXZYSXDG20}.                            \\
    \textbf{Alert Strategy}   & The policy of alert generation, including \textit{when to generate an alert}, \textit{what attributes and descriptions an alert should have}, and \textit{to whom the alert should be sent}.                    \\
    \textbf{SOP}              & A predefined Standard Operating Procedure (SOP) to inspect the state of the cloud system and mitigate the system abnormality upon receiving an alert. The operations can be conducted by OCEs or automatically. \\
    \textbf{Alert Governance} & The unified management of \textit{alert strategies} and \textit{SOPs}.                                                                                                                                          \\
    \bottomrule
  \end{tabularx}
  \vspace{-1em}
\end{table*}

In industrial practice, a cloud provider usually deploys a cloud monitoring system to obtain the telemetry data that reflects the running state of their cloud services~\cite{DBLP:conf/nsdi/LiCHDHSYLWLC20, theory-of-monitoring}.
Multiple monitoring techniques are employed to collect various types of telemetry data, including the performance indicators of the monitored service, the low-level resource utilization, the logs printed by the monitored service, etc.
For normally functioning services, it is assumed that their states, as well as their telemetry data, will be stable.
For a service that will fail soon, its telemetry data will fluctuate from the normal state~\cite{DBLP:conf/hotos/HuangGZLDCY17,DBLP:journals/corr/abs-2108-00344}.
Hence, cloud providers typically conduct anomaly detection on the telemetry data to detect the deviation from the normal state.
If an anomaly triggers an alert strategy, an alert will be generated, and the cloud monitoring system will notify OCEs according to the configuration of the alert strategy.

The configuration of alert strategies is empirical, which heavily depends on human expertise.
Since different cloud services exhibit different attributes and serve different purposes, their alert strategies vary significantly.
In particular, the empiricalness of alert strategies results from two aspects of cloud services.
On the one hand, a cloud service's abnormal state may differ because each cloud service implements its own business logic.
There is no one-fits-all rule for anomaly detection on cloud services, i.e., \textit{when to generate an alert}.
For example, network overload is a crucial anomaly for a virtual network service.
However, high connection number becomes a real issue for a database service.
On the other hand, the attributes of an alert that helps the manual inspection and mitigation of the abnormal state, e.g., the location information and the free-text title that describes the alert, are also service-specific and lack comprehensive guidelines.
In other words, ``\textit{what attributes and descriptions an alert should have}'' also depends on human expertise.
For example, the title ``Instance \emph{x} is abnormal'' is non-informative.
In summary, the configuration of alert strategies, as a precursor step for human intervention in cloud anomalies, is an empirical procedure.

Manually-configured alert strategies are flexible but can also be ineffective (e.g., misleading, non-informative, and non-actionable) when the engineer is inexperienced or unfamiliar with the monitored cloud service.
The ineffectiveness of alerts becomes anti-patterns that hinder the OCEs' diagnosis, especially for inexperienced OCEs.
The anti-patterns of alerts, which we will elaborate in Section~\ref{sec:empirical}, will frustrate OCEs and deteriorate cloud reliability in the long term.

In this paper, we conduct the first empirical study on the industrial practice of alert governance in \company\footnote{\company is a global cloud provider and ranked fifth in Gartner's report~\cite{gartner-market-share} on the global market share of \textit{Infrastructure as a Service} in 2020.}.
The cloud system considered in this study consists of 11 cloud services and 192 cloud microservices.
The procedure of our study includes 1) a quantitative assessment of over 4 million alerts in the time range of two years to identify the anti-patterns of alerts; 2) interviews with 18 experienced on-call engineers (OCEs) to confirm the identified anti-patterns and summarize the current practice to mitigate the identified anti-patterns.
To sum up, we make the following contributions:

\begin{itemize}[leftmargin=*, topsep=0pt]
  \item We conduct the first empirical study on characterizing and mitigating anti-patterns of alerts in an industrial cloud system.
  \item We identify six anti-patterns of alerts in a production cloud system.
        Specifically, the six anti-patterns can be divided into two categories, namely individual anti-patterns and collective anti-patterns.
        Individual anti-patterns result from the ineffective patterns in one single alert strategy, including \emph{Unclear Name or Description}, \emph{Misleading Severity}, \emph{Improper and Outdated Alert Strategy}, and \emph{Transient and Toggling Alerts}.
        Collective anti-patterns are ineffective patterns that a bunch of alerts collectively exhibit, including \emph{repeating} and \emph{cascading alerts}.
  \item We summarize the current industrial practices for mitigating the anti-patterns of alerts, including postmortem reactions to mitigate the effect of anti-patterns and the preventative guidelines to avoid the anti-patterns.
        The postmortem reactions include \emph{rule-based alert blocking} and \emph{alert aggregation}, \emph{pattern-based alert correlation analysis}, and \emph{emerging alert detection}.
        We also describe three aspects of designing preventative guidelines for alert strategies according to our experience in \company.
  \item Lastly, we share our thoughts on prospective directions to achieve automatic alert governance. We propose to bridge the gap between manual alert strategies and cloud service upgrades by automatically evaluating the Quality of Alerts (\qoa) in terms of \emph{indicativeness}, \emph{impact}, and \emph{handleability}.
\end{itemize}

\section{Alerts for the Reliability of Cloud Services}
\label{sec:preliminary}

This section provides the preliminary knowledge for our study.
We first generally introduce the reliability measures of cloud services, then describe the mechanism of alerting in cloud systems.

\subsection{Reliability of Cloud Services}
\label{sec:preliminary:architecture}

Cloud providers typically split various services into microservices and organize them into microservice architecture~\cite{berkeley-view-cloud}.
Microservices are small, independent, and loosely coupled modules that can be deployed independently~\cite{microservice-architecture}.
Communicating through well-defined APIs, each microservice can be refactored and scaled independently and dynamically~\cite{villamizar2015evaluating}.
External requests are routed through and served by dozens of different microservices that rely on one another.

%~ hard to diagnose
One of the major weaknesses of the microservice architecture is the difficulty in system maintenance~\cite{DBLP:conf/icse/Zhao0PWWZCZNWWZ20, DBLP:conf/sigsoft/Zhou0X0JLXH19}.
The highly decoupled nature of the microservice architecture makes the performance debugging, failure diagnosis, and fault localization in cloud systems more complex than ever~\cite{DBLP:conf/sigsoft/ZhangXQHQLZLDLC21, DBLP:conf/sigsoft/ChenKLZZXZYSXDG20, DBLP:conf/usenix/ZhangLXQ0QDYCCW19, DBLP:conf/asplos/GanZHCHPD19}.
A common pathway to tackle the difficulties in system maintenance is to 1) improve system observability~\cite{DBLP:conf/osdi/HuangGLZD18, log-survey, DBLP:conf/icse/PecchiaCCC15, DBLP:journals/ese/YaoPSSTS20, DBLP:conf/sigsoft/HeLLZLZ18} with logging, tracing, and performance monitoring, 2) employ proper alert strategies to detect system anomalies and send alerts~\cite{DBLP:conf/icse/Zhao0PWWZCZNWWZ20}, and 3) design effective SOPs to quickly mitigate the system abnormality before it escalates to severe failure and incidents.
In practice, cloud providers usually deploy cloud monitoring systems to improve observability, detect anomalies, and generate alerts.

\subsection{Alerts in Cloud Services}
\label{sec:preliminary:alert}

\begin{table*}[htbp]
    \centering
    \small % font size
    \caption{Sample reliability alerts in a cloud system. The names of microservices are omitted due to confidentiality.}
    \label{tab:sample-alerts}
    \begin{tabularx}{2\columnwidth}{llllXll}
        \toprule
        No. & Severity & Time             & Service       & Alert Title                              & Duration & Location          \\
        \midrule
        1   & Major    & 2021/05/18 06:36 & Block Storage & Failed to allocate new blocks, disk full & 10 min   & Region=X;DC=1;... \\
        2   & Critical & 2021/05/18 06:38 & Database      & Failed to commit changes ...             & 2 min    & Region=X;DC=1;... \\
        3   & Critical & 2021/05/18 06:39 & Database      & Failed to commit changes ...             & 5 min    & Region=X;DC=1;... \\
        \bottomrule
    \end{tabularx}
    \vspace{-1em}
\end{table*}

\subsubsection{Necessities of Alerts}
\label{sec:preliminary:alert:necessities}

\begin{figure}[htbp]
    \centering
    \includegraphics[width=\columnwidth]{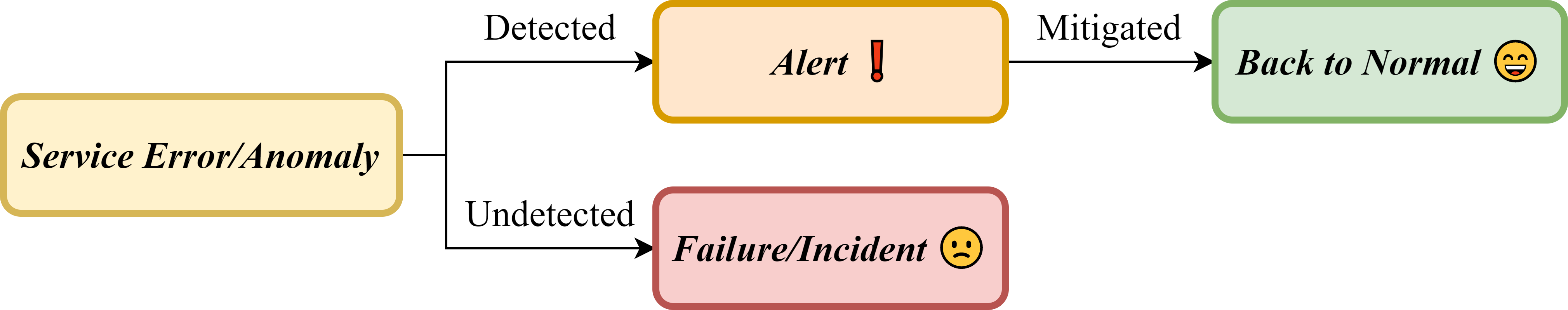}
    \vspace{-1.5em}
    \caption{The significance of alerts for cloud reliability.}
    \label{fig:alert-necessity}
    \vspace{-0.5em}
\end{figure}

% anomaly -> failure -> detect anomaly before failure
Service reliability is one of the most significant factors for cloud providers and their clients, but failures that prevent cloud services from properly functioning are inevitable~\cite{DBLP:conf/sigsoft/ChenKLZZXZYSXDG20}.
In order to satisfy Service Level Agreements (SLAs) on the reliability of the target services, cloud providers need to deal with service and microservice anomalies before they escalate their effect into severe failures and incidents.
Alerting is a practical way to achieve this goal.
Figure~\ref{fig:alert-necessity} demonstrates the significance of alerts.
By continuously monitoring cloud services via traces, logs, metrics, the monitoring system will send alerts\footnote{This paper only focuses on the alerts that indicate potential bugs and failures, i.e., the system reliability alerts.} to On-Call Engineers (OCEs) upon detecting anomalous service states.
With the information provided in the alerts, OCEs can judge with their domain knowledge, fix the problems, and clear the alert.
As a result, unplanned failures and incidents can be avoided or quickly mitigated.

\vspace{-0.5em}

\subsubsection{Attributes of Alerts}
\label{sec:preliminary:alert:attributes}

Alerts have many attributes that are helpful for OCEs' diagnosis, including title of alerts, severity level, time, service name, duration, location information.
The \emph{Title of an Alert} concisely describes the alert.
Typically, the title should contain information like ``the affected service or microservice'' and ``the manifestation of the failure''.
The OCEs will look up the alert title to find the corresponding SOP and perform predefined actions to mitigate the alert.
The \emph{Severity Level} indicates how severe the alert is.
The corresponding \emph{Alert Strategy} defines the severity level and alert title according to the nature of the affected service or microservice.
The \emph{Time} means the time of occurrence of the alert, and \emph{Duration} is the duration between the occurrence and the clearance of the alert.
The \emph{Location Information} contains the necessary information to locate the anomalous service or microservice.
Table~\ref{tab:sample-alerts} shows the samples of alerts from the monitoring system of \company.

\subsubsection{Generation of Alerts}
\label{sec:preliminary:alert:generation}

An alert represents a specific abnormal state of the cloud system.
The first and foremost step of alert generation is anomaly detection.
Anomaly detection in logs~\cite{log-survey,DBLP:journals/corr/abs-2108-01955,DBLP:conf/sigsoft/ZhaoWLPWPWFWZSP21}, traces~\cite{AID, DBLP:conf/sigsoft/Guo0WLJDXS20, DBLP:conf/sigsoft/Zhou0X0JLXH19}, and monitoring metrics~\cite{DBLP:conf/iwqos/MengZSZHZJWP20, DBLP:conf/issre/LiuCNZZSZP19, DBLP:journals/tosem/ZhaoHZTC21} of the cloud system have been widely studied.

The cloud monitoring system will continuously detect anomalies and generate system reliability alerts according to the alert strategies associated with specific services or microservices.
The strategies for system reliability alerts can be divided into three categories, i.e., probes, logs, and metrics.

\begin{itemize}[leftmargin=*] %~ noindent
    \item \emph{Probes:}
          The cloud monitoring system will send probing requests to the target services and receive the heartbeat from the target services.
          Typically, OCEs set fixed thresholds of no-response time for different services as the strategy of probes.
          If a target service does not respond to the probing requests for a long time, an alert will be generated.
    \item \emph{Logs:}
          The cloud monitoring system will process logs of the target services.
          OCEs can set flexible rules for different services.
          Typical rules of logs are keyword matching, e.g., ``IF the logs contain 5 \texttt{ERROR}s in the past 2 minutes, THEN generate an alert.''
          Traces can also be viewed as special logs and will be processed similarly.
    \item \emph{Metrics:}
          Performance metrics are time series that show the states of a running service, e.g., latency, no. of requests, network throughput, CPU utilization, disk usage, memory utilization, etc.
          The alert strategy for metrics varies from static threshold to algorithmic anomaly detection.
\end{itemize}

\subsubsection{Clearance of Alerts}
\label{sec:preliminary:alert:clearance}

Alerts can be cleared manually or automatically.
On the one hand, after the human intervention, if the OCE confirms the mitigation of the anomaly, the OCE can manually mark the alert as ``cleared''.
On the other hand, the cloud monitoring system can automatically clear some alerts.
For system reliability alerts of \emph{probes} and \emph{metrics}, the cloud monitoring system will continue to monitor the status of the associated service.
If the service returns to a normal state, the cloud monitoring system will mark the corresponding alert as ``automatically cleared''.

\section{An Empirical Study on the Anti-patterns of Alerts}
\label{sec:empirical}

The research described in this paper is motivated by the pain point of alert governance in a production cloud system.
In this section, we present the first empirical study of characterizing the anti-patterns of alerts\footnote{An alert always corresponds to an alert strategy. Therefore, we do not discriminate ``anti-pattern of alerts'' and ``anti-patterns of alert strategies''.} and how we mitigate the anti-patterns in the production cloud system.
Specifically, we study the following research questions (RQs).

\begin{itemize}[leftmargin=*, topsep=0pt]
  \item \textbf{RQ1}: What anti-patterns exist in alerts? How do these anti-patterns prevent OCEs from promptly and precisely diagnosing the alert?
  \item \textbf{RQ2}: What is the standard procedure to process alerts? Can the standard procedure handle the anti-patterns?
  \item \textbf{RQ3}: What are the current \textbf{reactions} to the anti-patterns of alerts? How about their performance?
  \item \textbf{RQ4}: What are the current measures to \textbf{avoid} the anti-patterns of alerts? How about their performance?
\end{itemize}

To answer these research questions, we quantitatively analyzed over 4 million alerts from the production system of \company which serves tens of millions of users and contains hundreds of services.
The time range of the alerts spans over two years.
We conducted a survey involving 18 experienced OCEs to find out the current practice of mitigating the anti-patterns of alerts.
Among them, 10 (55.6\%) OCEs have more than 3 years of working experience.
The number of OCEs with 2 to 3 years' working experience and 1 to 2 years' working experience are 3 (16.7\%) and 2 (11.1\%).
Lastly, 3 (16.7\%) OCEs' experience are less than 1 year.

\begin{figure*}[!t]
  \centering
  % \subfigure[How long have you been working as an OCE?]{
  %   \includegraphics[width=0.32\columnwidth]{figures/survey-experience.pdf}%
  %   \label{fig:survey-experience}
  % }
  % \hfil
  \subfigure[How about the impact of different anti-patterns to alert diagnosis?]{
    \includegraphics[width=0.55\columnwidth]{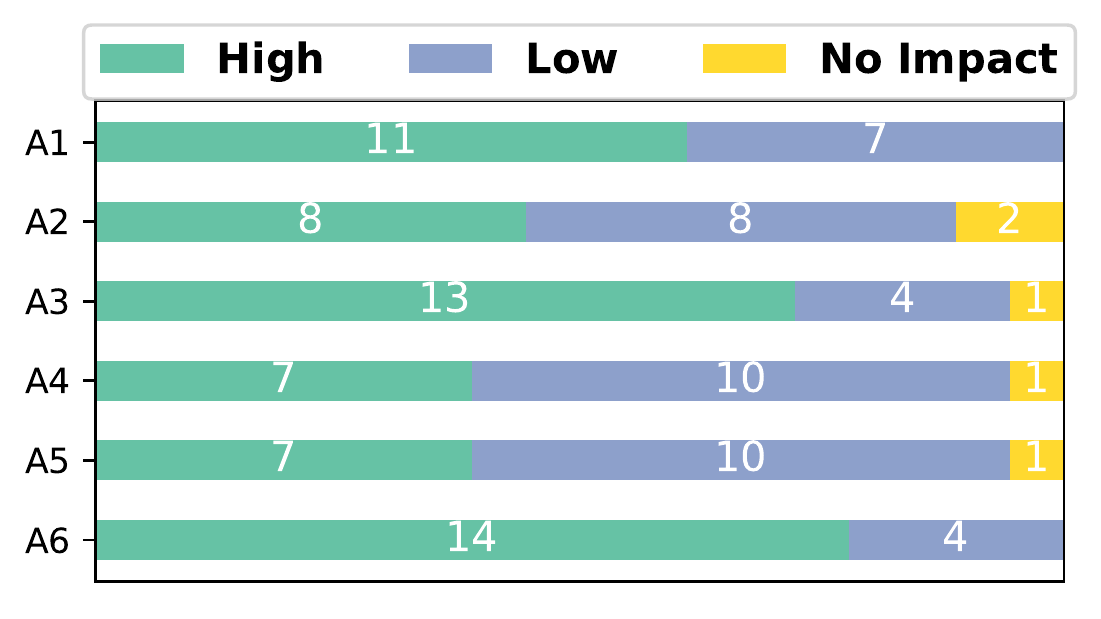}%
    \label{fig:survey-impact}
  }
  % \hfil
  \subfigure[How helpful are the predefined SOPs?]{
    \includegraphics[width=0.67\columnwidth]{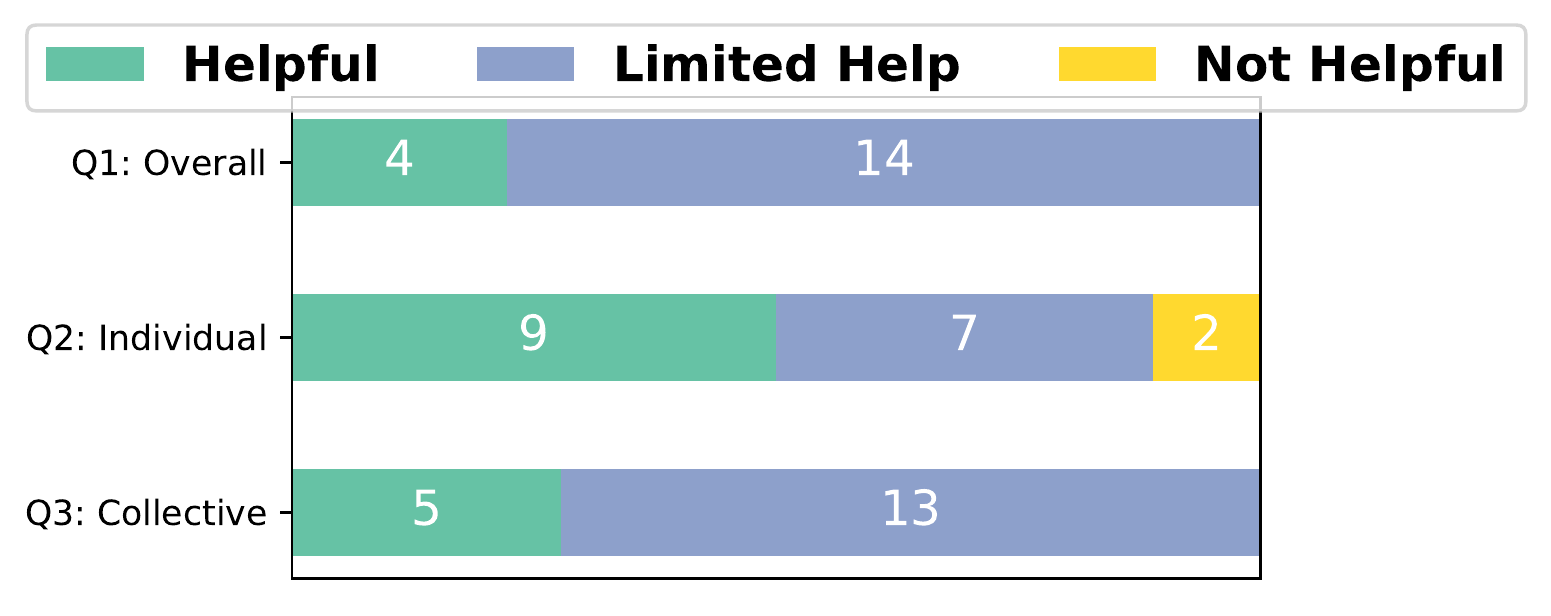}%
    \label{fig:survey-sop-helpfulness}
  }
  % \hfil
  \subfigure[How about the effectiveness of current reactions to anti-patterns?]{
    \includegraphics[width=0.67\columnwidth]{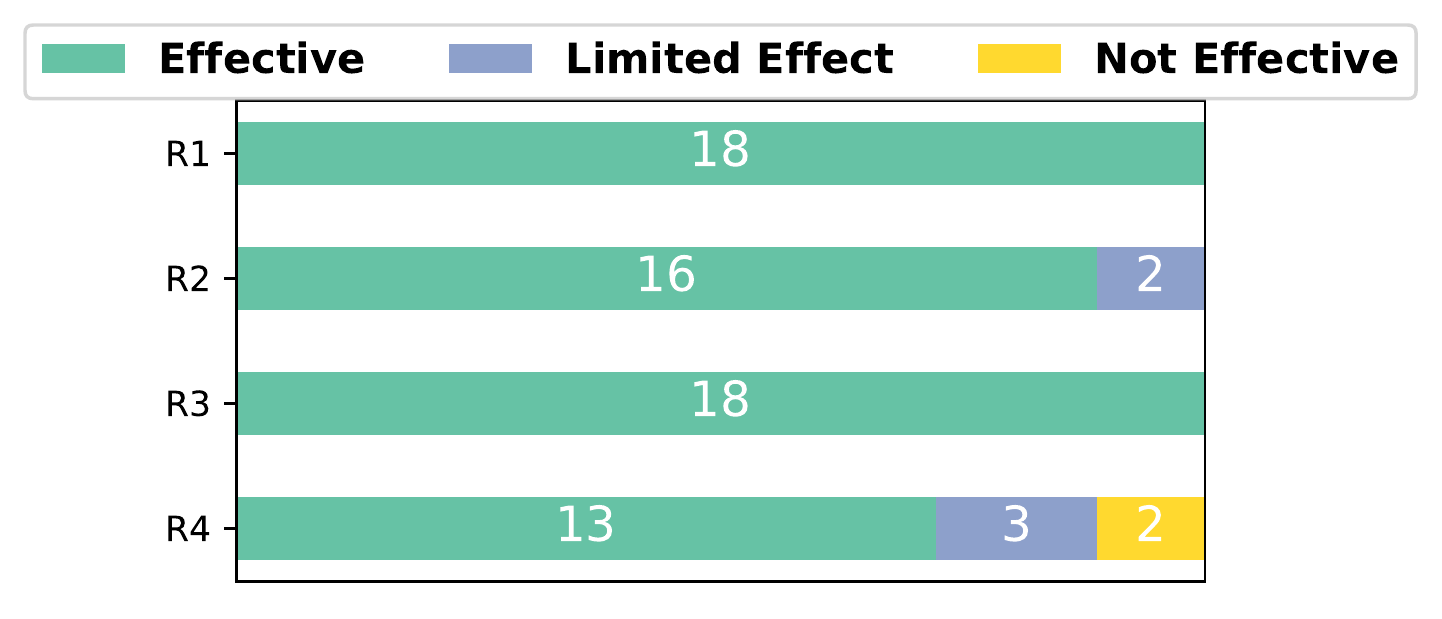}%
    \label{fig:survey-reaction-helpfulness}
  }
  \vspace{-0.5em}
  \caption{A survey about the current practice of mitigating the anti-patterns of alerts.}
  \vspace{-0.5em}
  \label{fig:survey}
  \vspace{-1em}
\end{figure*}

\subsection{RQ1: Anti-patterns in Alerts}
\label{sec:empirical:anti-patterns}

Anti-patterns of alerts are misconfigured and ineffective patterns in alerts that hinder alert processing in practice.
Although alerts provide essential information to OCEs for diagnosing and mitigating failures, anti-patterns of alerts hinder this process.
We divide the anti-patterns into two categories, i.e., \textit{individual anti-patterns} and \textit{collective anti-patterns}.
\textit{Individual anti-patterns} result from the ineffectiveness of one single alert.
In practice, OCEs usually have limited time to diagnose alerts.
If one alert and its SOP are poorly designed, e.g., misleading steps to diagnose or non-informative description, the manual diagnosis will be difficult.
\textit{Collective anti-patterns} are ineffectiveness that alerts collectively exhibit.
Sometimes, due to inappropriate configuration of alert strategy, complex dependency, and inter-influence effect in the cloud, numerous alerts may simultaneously occur.
If alerts flood to OCEs or are collectively hard to handle, it will be too complicated for manual diagnosis, especially for inexperienced OCEs.
Characterizing these anti-patterns is the leading step for alert governance.

For this research question, we analyzed more than 4 million alerts over two years to characterize the anti-patterns of alerts.
The total number of alert strategies in this empirical study is $2010$.
To select the candidates of individual anti-patterns, we group the alerts according to the alert strategies, then calculate each strategies' average processing time.
The alert strategies that take the top 30\% longest time to process are selected as the candidates of individual anti-patterns.
To find cases of collective anti-patterns, we first group all the alerts by the hour they occur and the region they belong to.
Then we count the number of alerts per hour per region.
If the number of alerts per hour per region exceeds 200\footnote{We set the threshold as 200 as the estimated maximum number of alerts an OCE team can deal with is 200. Experienced OCEs confirm the threshold.}, we select all the alerts in this group as the candidate of collective anti-patterns.
We also went through the incident reports over the past two years to seek the ineffectiveness in alerts recorded by OCEs.
We get five candidate cases of individual anti-patterns and two candidate cases of collective anti-patterns.
After that, we ask two experienced OCEs to mark whether they think the candidate ineffective pattern in alerts is an anti-pattern.
If they both agree, we include it as an anti-pattern.
If disagreements occur, another experienced OCE is invited to examine the pattern.
As a result, we summarized four individual anti-patterns and two collective anti-patterns.

Our survey asked the OCEs to determine the impact of different anti-patterns on alert diagnosis.
Figure~\ref{fig:survey-impact} shows the answers' distributions.
Each bar represents one anti-pattern, which is elaborated below.

\subsubsection{Individual anti-patterns}%

Individual anti-patterns are the ineffectiveness of a single alert, including unclear name or description, misleading severity, and improper and outdated generation rule.

\textbf{[A1]} \emph{Unclear Name or Description}.
Unclear alert name or alert description obstructs the OCEs from gaining intuitive judgment at the first sight, which slows down the diagnosis and even hinders OCEs from knowing the logical connections from the alert to other alerts.
Typical unclear alert names describe the system state in a very general way with vague words, e.g., ``Elastic Computing Service is abnormal'', ``Instance $x$ is abnormal'', ``Component $y$ encounters exceptions'', and ``Computing cluster has risks''.
All OCEs agree with the impact of \emph{unclear name or description}, and 61.1\% of them think the impact is high.

\textbf{[A2]} \emph{Misleading Severity}.
Severity helps OCEs to prioritize which alert to diagnose first.
Inappropriately high severity level takes up OCE's time for dealing with less essential alerts, while too low severity level may lead to missing important alerts.
In our survey, 88.9\% of OCEs agree with the impact of \emph{misleading severity}.
In practice, we find that the setting of severity heavily depends on domain knowledge.
With the update of the cloud system, especially the enhancement of fault tolerance mechanisms, the severity may also change.

\textbf{[A3]} \emph{Improper and Outdated Generation Rule}.
Typically, the cloud monitoring system will continuously monitor the performance indicators of both lower-level infrastructures (e.g., CPU usage, disk usage) and higher-level services (e.g., request per second, response latency).
If any indicator increases over or drops below the predefined thresholds, an alert will be generated.
Although the performance indicators of lower-level infrastructures can provide valuable information when the root cause of the alert is failures of lower-level infrastructures (e.g., high CPU usage), due to the fault-tolerance techniques applied in cloud services, the performance indicators of lower-level infrastructures do not have definite effect on the quality of cloud services from the perspective of customers.
According to our survey, 72.2\% of OCEs agree that the impact of \emph{improper and outdated generation rule} is high.

\textbf{[A4]} \emph{Transient and Toggling Alerts}.
As mentioned in Section~\ref{sec:preliminary:alert:clearance}, the cloud monitoring system can automatically clear some alerts.
When the interval between the generation time and automatic clearance time of an alarm is less than a certain value (known as the intermittent interruption threshold), the alert is called a transient alert.
Commonly speaking, a transient alert is an alert that lasts for a short time.
When the same alert is generated and cleared multiple times (i.e., oscillation), and the number of oscillations is greater than a certain value (known as the oscillation threshold), it is called a toggling alert.
Transient and toggling alerts are usually caused by alert strategies being too sensitive to the fluctuation of the metrics.
Transient and toggling alerts cause fatigue of OCEs and also distract the OCEs from being dealing with other important alerts.
Although there are disagreements on the level of impact, most OCEs (94.4\%) think the impact exists.

\subsubsection{Collective anti-patterns}%

Collective anti-patterns result from the ineffective patterns of a bunch of alerts that occur in a short time scope.
Zhao et al.~\cite{DBLP:conf/icse/Zhao0PWWZCZNWWZ20} defined numerous alerts (e.g., hundreds of alerts) from different cloud services in a short time (e.g., one minute) as ``alert storm'', and conducted several case studies of alert storms.
In alert storms, even if all the individual alerts are effective, the large number of alerts may still set obstacles for OCEs and greatly affect the system reliability in the following three ways.
Firstly, during an alert storm, many alerts are generated.
If OCEs check each alert manually, the troubleshooting will take unacceptably long time.
Secondly, since alert storms occur frequently~\cite{DBLP:conf/icse/Zhao0PWWZCZNWWZ20}, the OCEs will continually receive alerts by email, SMS, or even phone call.
According to our study, alert storms occur weekly or even daily, and 17 out of 18 interviewed OCEs say that the alert storms greatly fatigue them.
Lastly, the overwhelming number of alerts adds pressure to the monitoring system, so the latency of generating new alerts may increase.

Inspired by~\cite{DBLP:conf/icse/Zhao0PWWZCZNWWZ20}, we summarize the following collective anti-patterns from confirmed cases of alert storms in \company.
In this study, if the number of alerts from a region exceeds 100 in an hour, we count it as an alert storm.
Consecutive hours of alert storm will be merged into one.
Among the two collective anti-patterns, ``cascading alerts'' has already been observed by~\cite{DBLP:conf/icse/Zhao0PWWZCZNWWZ20}, but ``repeating alerts'' has not.
In particular, we demonstrate the collective anti-patterns of alerts with a representative alert storm that happened from 7:00 AM to 11:59 AM in \company.
During the alert storm, totally 2751 alerts were generated, among which we observeed both collective anti-patterns as described below.

\vspace{-0.5em}
\begin{figure}[htbp]
  \centering
  \begin{minipage}[t]{0.57\columnwidth}
    \centering
    \includegraphics[width=\columnwidth]{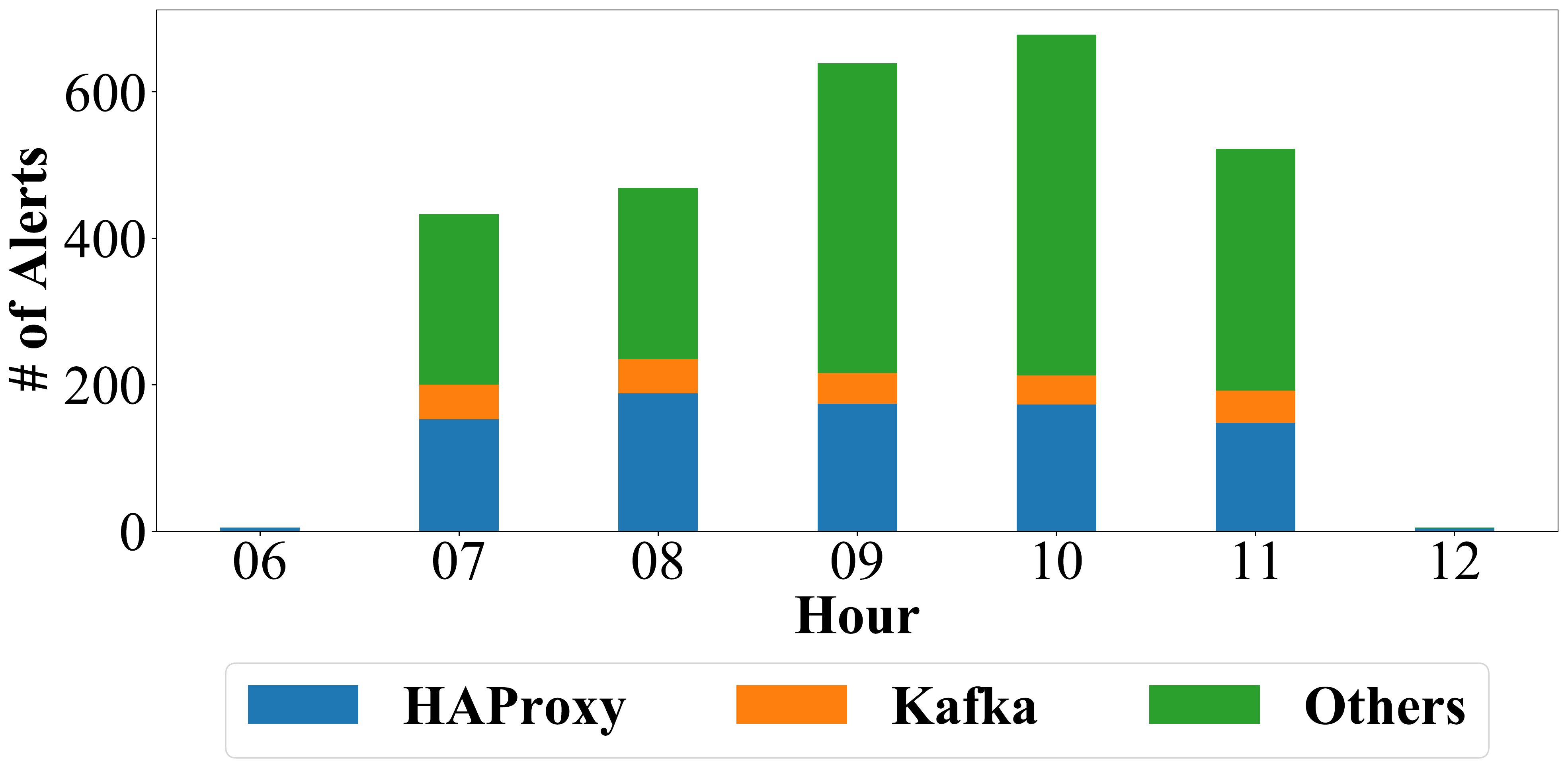}
    \label{fig:repeating-alert}
    \vspace{-1.5em}
    \caption{Repeating alerts in an alert storm.}
  \end{minipage}
  \hfill
  \begin{minipage}[t]{0.41\columnwidth}
    \centering
    \includegraphics[width=\columnwidth]{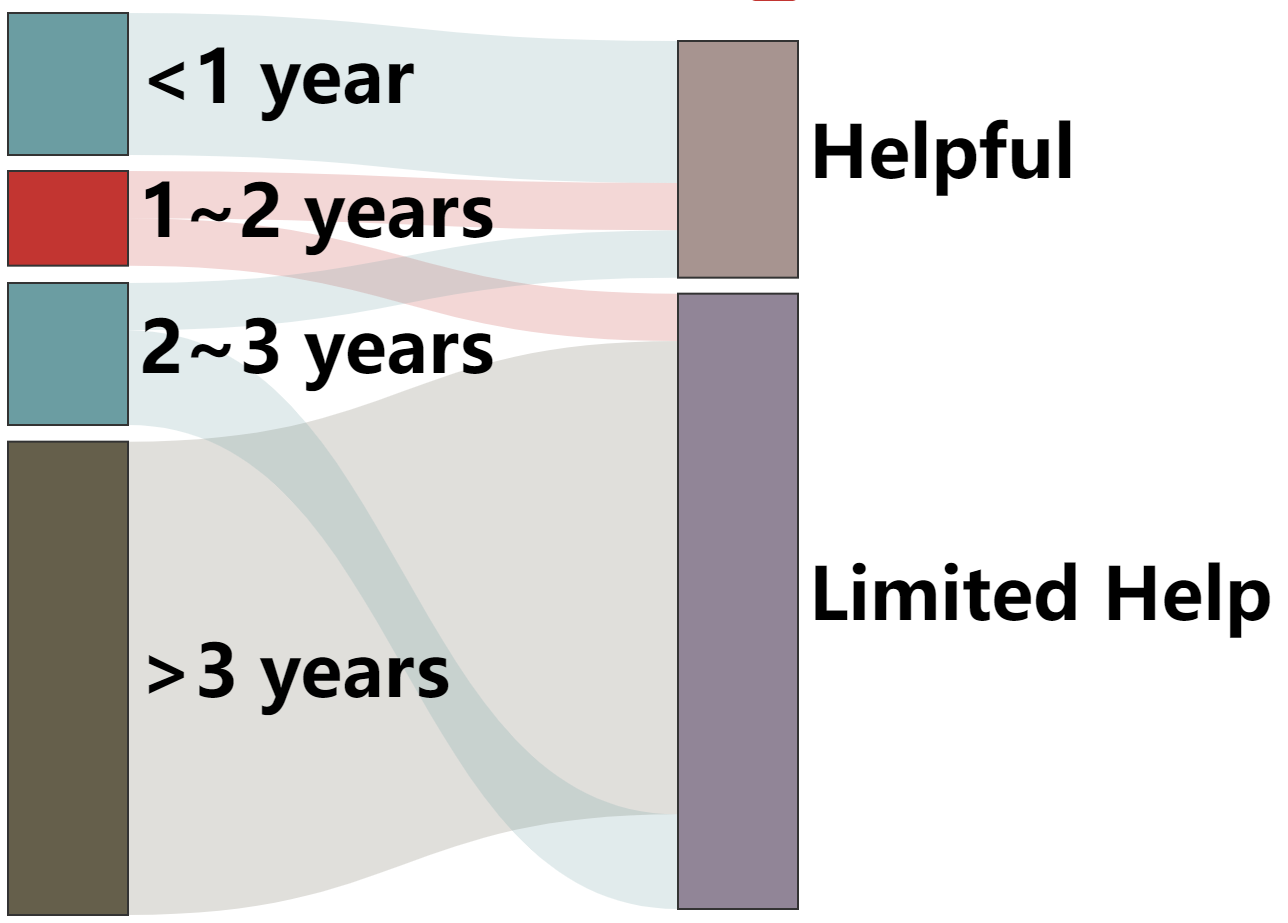}
    \label{fig:sankey}
    \vspace{-1.5em}
    \caption{Answers to Q1 ``Overall Helpfulness'' regarding OCEs' working experience.}
  \end{minipage}
\end{figure}
\vspace{-0.5em}

\textbf{[A5]} \emph{Repeating Alerts}.
Repeating alerts means that alerts from the same alert strategy appear repeatedly.
Sometimes the repeated alerts may last for several hours.
This is usually due to the inappropriate frequency of alert generation.
For example, in Figure~\ref{fig:repeating-alert}, we count the number of alerts per strategy.
The total number of alerts is 2751, and the number of effective alert strategies is 200.
To make the figure clear, we only show the name of the top two alerts.
All other alerts are classified as ``Others'' in the figure.
The alert ``haproxy process number warning'', abbreviated as \texttt{HAProxy} in the figure, takes up around 30\% of the total number of alerts in each hour.
However, it is only a \texttt{WARNING} level alert, i.e., the lowest level.
Even though an individual alert is straightforward to process, it is still time-consuming to deal with it when it occurs repeatedly.
If one rule continually generates alerts, it will distract OCEs from dealing with the more essential alerts.
Most OCEs (94.4\%) agree with the impact of \emph{repeating alerts}.

\textbf{[A6]} \emph{Cascading Alerts}.
Modern cloud systems are composed of many microservices that depend on each other~\cite{AID}.
When a service enters an anomalous state, other services that rely on it will probably suffer from anomalous states as well.
Such anomalous states can propagate through the service-calling structure~\cite{aws-architecture}.
Despite various fault tolerance mechanisms being introduced, minor anomalies are still common to magnify their impact and eventually affect the entire system.
Each of the affected services will generate many anomalous monitoring metrics, resulting in many alerts (e.g., thousands of alerts per hour).
As a consequence, the alerts burst and flood to the OCEs.
Although the alerts are different, they are implicitly related because they originate from the cascading effect of one single failure.
Manually inspecting the alerts is hard without sufficient knowledge of the dependencies in the cloud system.
All interviewed OCEs agree with the impact of \emph{cascading alerts}.
Table~\ref{tab:sample-alerts} shows a simplified sample of cascading alerts.
By manually inspecting the alerts, experienced OCEs would infer that the alert 1 possibly cause alert 2 because 1) Alert 2\&3 occurred right after alert 1 and 2) The relational database service relies on the block storage service as the backend.
If the relational database service failed to commit changes, i.e., write data, one possible reason is that the storage service failed.

\begin{mybox}
  \textbf{Finding 1}: Individual anti-patterns and collective anti-patterns widely exist. They hinder alert diagnosis to different extent.
\end{mybox}

\subsection{RQ2: Standard Alert Processing Procedure}
\label{sec:empirical:general-procedure}

\vspace{-1em}
\begin{figure}[htbp]
  \centering
  \includegraphics[width=0.9\columnwidth]{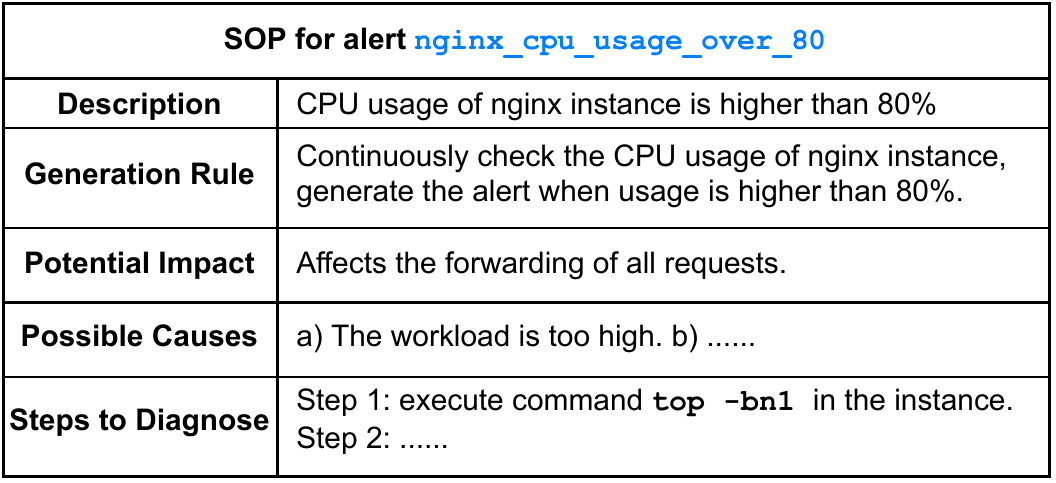}
  \vspace{-0.5em}
  \caption{An example Standard Operation Procedure.}
  \vspace{-0.5em}
  \label{fig:sop}
\end{figure}

The Standard Operation Procedure (SOP) defines the procedure to process a single alert.
For each alert, its SOP includes the alert name, the alert description, the generation rule of the alert (i.e., alert strategy), the potential impact on the cloud system, the possible causes, and the steps to process the alert.
Figure~\ref{fig:sop} shows an example SOP of the alert \texttt{nginx\_cpu\_usage\_over\_80}.
The OCEs can follow the SOP to process the alert upon receiving the alert.
According to our survey, only 22.2\% of OCEs think current SOPs are helpful (Q1, Figure~\ref{fig:survey-sop-helpfulness}), and the other 77.8\% of OCEs say the help is limited.
The SOPs are deemed to show limited help by all OCEs with over 3 years' experience, taking up $71.4\%$ of all OCEs selected "Limited Help" for Q1 (Figure~\ref{fig:sankey}).
Moreover, SOPs are considered much less helpful for diagnosing collective anti-patterns (Q3, Figure~\ref{fig:survey-sop-helpfulness}) than individual anti-patterns (Q2, Figure~\ref{fig:survey-sop-helpfulness}).

\begin{mybox}
  \textbf{Finding 2}: SOPs can help OCEs quickly process alerts, but the help is limited. SOPs are considered less helpful when dealing with collective anti-patterns.
\end{mybox}

\subsection{RQ3: Reactions to Anti-patterns}
\label{sec:empirical:reaction}

Depending on the number of alerts, OCEs react differently.
When the number of alerts is relatively small, OCEs will scan through all the reported alerts.
Then they will manually rule out alerts that are not of great importance and deal with critical alerts that will affect the whole system.

OCEs react differently when the number of alerts becomes too large.
According to our interview with senior OCEs in \company, they typically take four kinds of reactions, i.e., alert blocking, alert aggregation, alert correlation analysis, and emerging alert detection.
In practice, we observe that although the reactions are considered effective, they need to be reconfigured after the update of cloud services or alert strategies.

\textbf{[R1]} \emph{Alert Blocking}.
When OCEs find that transient alerts, toggling alerts, and repeating alerts provide no information about service anomaly, they can treat these alerts as noise and block them with alert blocking rules.
As a result, these non-informative alerts will not distract OCEs from quickly identifying the root causes of service anomalies.

\textbf{[R2]} \emph{Alert Aggregation}.
When dealing with large amounts of alerts, there may be many duplicate alerts in a time period.
For the non-informative alerts, OCEs will employ alert blocking introduced before to facilitate analysis.
For the informative ones, they will adopt alert aggregation.
To be more specific, OCEs will set rules to aggregate alerts in a period and use the number of alerts as another feature~\cite{DBLP:journals/corr/abs-2108-12179}.
By doing so, OCEs can quickly identify critical alerts and focus more on the information provided by them.

\textbf{[R3]} \emph{Alert Correlation Analysis}.
Apart from the information provided by the alerts and their statistical characteristics, OCEs will also leverage other exogenous information to analyze the correlation of alerts.
Two kinds of exogenous information are used to correlate alerts.
The first is the dependencies of alert strategies, which indicate the spread of alerts in the cloud services~\cite{DBLP:conf/wetice/MeloM19}.
For instance, if a source alert triggers another alert, OCEs will be more interested in the source alert, potentially the root cause of future service failures.
They will associate all the derived alerts with their source alerts and diagnose the source alerts only.
Another exogenous information is the topology of cloud services.
Based on the topology of services, OCEs will set rules to correlate alerts based on the services that generated them.
With this kind of correlation, OCEs can quickly pinpoint the root cause of a large number of alerts by following the topological correlation.

\textbf{[R4]} \emph{Emerging Alert Detection}.
Due to the large scale of cloud services, manually configured dependencies of alert strategies could not cover all the alert strategies.
This may lead to the failure of alert correlation analysis.
For example, a few alerts corresponding to a root cause (i.e., emerging alerts) appear first.
If they are not dealt with seriously, when the root cause escalates its influence, numerous cascading alerts will be generated.
The lack of critical association rules will prevent the OCEs from discovering the correlation and quickly alert diagnosis.
This usually happens on gray failures like memory leak and CPU overloading.
Hence, it would be helpful to capture the implicit dependencies.
We employ the adaptive online Latent Dirichlet Allocation~\cite{DBLP:conf/www/YangGZ0L21, DBLP:conf/icse/GaoZLK18} to capture the implicit dependencies.
OCEs could detect these emerging alerts as early as possible for faster alert diagnosis with the implicit dependencies.

Figure~\ref{fig:survey-reaction-helpfulness} shows OCEs' opinions about the effectiveness of the four reactions.
In general, the effectiveness of all four reactions is relatively high.

\begin{mybox}
  \textbf{Finding 3}: Current reactions are considered effective, but the configurations of such reactions still require domain knowledge.
\end{mybox}

\subsection{RQ4: Avoidance of Anti-patterns}
\label{sec:empirical:preventative}

To avoid the alert anti-patterns from occurring, \company also adopts preventative guidelines and conducts periodical reviews on alert strategies.
We summarize the generic aspects to consider when designing the guidelines.
The guidelines are designed by experienced OCEs and guide from three aspects of alerts.

\begin{itemize}[leftmargin=*, topsep=0pt]
  \item \emph{Target}
        means what to monitor.
        The performance metrics highly related to the service quality should be monitored.
  \item \emph{Timing}
        means when to generate an alert upon the manifestation of anomalies.
        Sometimes an anomaly does not necessarily mean the service quality will be affected.
  \item \emph{Presentation}
        means whether the alerts' attributes are helpful for alert diagnosis.
\end{itemize}

However, our interview with OCEs shows that the preventative guidelines are not strictly obeyed in practice.
Most (88.9\%) OCEs agree that strictly following the guidelines will make alert diagnosis easier.

\begin{mybox}
  \textbf{Finding 4}: The preventative guidelines could reduce the anti-patterns and assist in alert diagnosis if they are carefully designed and strictly obeyed.
\end{mybox}

\section{Future Directions}
\label{sec:future}

Although several postmortem reactions and preventative guidelines are adopted (Section~\ref{sec:empirical}), according to our study, the problem of alert anti-patterns is still prevailing in industrial cloud monitoring systems because most current measures still require manual configuration.
As for the alert blocking, OCEs need to inspect each alert and set rules manually.
How to define the blocking rules and when to invalidate these rules become a crucial problem.
A similar problem also exists in alert correlation.
As for alert correlation analysis, OCEs also need to inspect alert generating rules and service topology documents apart from reading alerts, which incurs a considerable burden to OCEs.
Moreover, the effectiveness of the reactions also lacks clear criteria to evaluate.
OCEs can only estimate the effectiveness of the reactive measures by their feeling.
Therefore, outdated reactive measures is hard to detect.
As a result, the whole process of alert governance becomes time-consuming and laborious.

\begin{figure}[htbp]
    \centering
    \includegraphics[width=0.7\columnwidth]{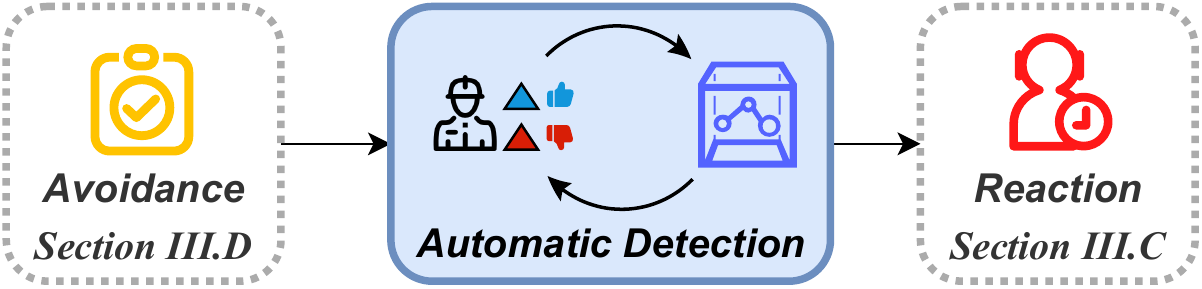}
    \vspace{-0.5em}
    \caption{Incorporating human knowledge and machine learning to detect anti-patterns of alerts.}
    \vspace{-0.5em}
    \label{fig:qoa}
\end{figure}

In Figure~\ref{fig:qoa}, we formulate the three stages of the mitigation of alert anti-patterns.
We already shared our experience of avoiding and reacting to alert anti-patterns in Section~\ref{sec:empirical}.
To close the gap between manual alert strategies and cloud system upgrades, we propose to explore the automatic detection of alert anti-patterns.
Automatic evaluation of the Quality of Alerts (\qoa) will be a promising approach to the automatic detection of alert anti-patterns.

Based on our empirical study, we propose three criteria to measure the quality of alerts (\qoa), including \textit{indicativeness}, \textit{precision}, and \textit{handleability}.

\begin{itemize}[leftmargin=*, topsep=0pt]
    \item \emph{Indicativeness} measures whether the alert can indicate the failures that will affect the end users' experience.
    \item \emph{Precision} measures whether the alert can correctly reflect the severity of the anomaly.
    \item \emph{Handleability} measures whether the alert can be quickly handled. The handleability depends on the target and the presentation of the alert. Improper target or unclear presentation decreases the handleability.
\end{itemize}

In the future, incorporating human knowledge and machine learning to evaluate the three aspects of alerts deserves more exploration.
In particular, OCEs provide their domain knowledge by creating labels like ``high/low precision/handleability/indicativeness'' for each alerts during alert processing. With the labels, a machine learning model could be trained and continuously updated so that it can automatically absorb the human knowledge for future \qoa evaluation.

\section{Related Work}
\label{sec:relatedwork}

Many works focus on processing alerts of cloud services and microservices.
One of the essential tasks of alert processing is to reduce the enormous amount of reported alerts to facilitate failure diagnosis.
Alert correlation~\cite{DBLP:conf/css/MirheidariAJ13} and clustering~\cite{DBLP:conf/kdd/LinRRYRF14,DBLP:conf/IEEEscc/XuWCW17,DBLP:conf/icse/Zhao0PWWZCZNWWZ20} are two common techniques employed to help OCEs find critical alerts and repair the system in a short period.
Li et al.~\cite{DBLP:conf/usenix/LiZZ0KZQHLSGYLR21} proposes to generate incidents based on the system alerts to prevent services from future failures.
Unlike all prior works, our paper focuses on not only how to deal with alerts after they are generated, but also how to generate better alerts and conduct better alert governance.

\section{Conclusion}
\label{sec:conclu}

This paper conducts the first empirical study to characterize the anti-patterns in cloud alerts.
We also summarize the industrial practices of mitigating the anti-patterns by postmortem reactions and preventative guidelines.
We wish our study to inspire further research on automatic \qoa evaluation and anti-pattern detection and benefit the reliability of the cloud services in the long run.

% acknowledge funds
\section*{Acknowledgment}

The work was supported by Key-Area Research and Development Program of Guangdong Province (No. 2020B010165002), Key Program of Fundamental Research from Shenzhen Science and Technology Innovation Commission (No. JCYJ20200109113403826), and the Research Grants Council of the Hong Kong Special Administrative Region, China (CUHK 14210920).



\bibliographystyle{IEEEtran}
\bibliography{main}

\end{document}